\documentstyle[aps,preprint]{revtex}

\begin{document}


\title{Dilaton gravity (with a Gauss-Bonnet term) derived    
from five-dimensional Chern-Simons gravity}

\author{M\'aximo Ba\~nados }

\address{Centro de Estudios
Cient\'{\i}ficos de Santiago, Casilla 16443, Santiago, Chile}

\maketitle

\begin{abstract}

We study the problem of boundary terms and boundary conditions for
Chern-Simons gravity in five dimensions.  We show that under
reasonable
boundary conditions one finds an effective field theory at the
four-dimensional boundary described by dilaton gravity with a 
Gauss-Bonnet term.  The coupling of matter is also discussed.   

\end{abstract}

The existing link between three-dimensional Chern-Simons theory and 
two-dimensional conformal field theory\cite{Witten89-etc} is a
remarkably powerful tool in the applications of Chern-Simons theory
to
2+1 gravity. Carlip \cite{Carlip95} has shown that the number of
states
of a conformal field theory lying at the horizon gives the correct
Bekenstein-Hawking expression for the 2+1 black hole entropy.  Also,
the
rich asymptotic structure of anti-de Sitter 2+1 gravity
\cite{Brown-Henneaux2} can be analyzed in a simple way as a WZW model
lying at the boundary  \cite{Henneaux-Coussaert-vanDriel,Ezawa,B}.

Chern-Simons theories exist in all odd-dimensional spacetimes. It is
therefore a natural question to ask whether they induce a `conformal'
field theory in a lower dimension. Steps in that direction were taken
in
\cite{Nair} where a generalization to four dimensions of the WZW
action
and its associated current algebra was constructed.  The $WZW_4$
theory
was also shown to be related to a Kahler-Chern-Simons theory.
Recently
\cite{BGH2}, the precise connection between pure Chern-Simons theory
and
the four-dimensional current algebra found in \cite{Nair} was
established  by studying the problem of boundary terms and boundary
conditions in Chern-Simons theory for groups of the form $G\times
U(1)$.

We consider in this paper the problem of boundary conditions and
boundary terms in a particular five-dimensional Chern-Simons theory
which is the analog of the Chern-Simons formulation of 2+1 gravity
studied in \cite{Achucarro-Townsend,Witten88}. We shall see that by 
imposing natural boundary conditions on the five-dimensional problem
one
obtains a four-dimensional field theory described by dilaton gravity
with a Gauss-Bonnet term. \\
 
\noindent{\em 1. Five-dimensional Poincare Chern-Simons gravity.}
We consider a five-dimensional Chern-Simons theory for the
group $ISO(3,2)$ [or $ISO(4,1)$] defined by the action  
\begin{equation}
I_{CS}=\frac{1}{2} \int_M \epsilon_{ABCDE} 
\tilde R^{AB}\mbox{\tiny $\wedge$} 
\tilde R^{CD}\mbox{\tiny $\wedge$} e^E ,
\label{I2}
\end{equation}
where $\tilde R^{AB}=dW^{AB}+W^A_{\ C} 
\mbox{\tiny $\wedge$} W^{CB}$ is the five-dimensional curvature
two-form. The fields $W^{AB}$ and $e^A$ can be collected together to
form an $ISO(3,2)$ [$ISO(4,1)$] connection, and (\ref{I2}) can be
shown to be an $ISO(3,2)$ [$ISO(4,1)$] Chern-Simons action  
\cite{Chamseddine,BTZ2,B-Tr-Z}. Indeed (\ref{I2}) is explicitly
invariant under $SO(3,2)$  [$SO(4,1)$] rotations. It is also
invariant,
up to a boundary term, under the Abelian translation $\delta e^A =
\nabla \lambda^A$, $\delta W^{AB}=0$. Hence, (\ref{I2}) is invariant
under $ISO(3,2)$  [$ISO(4,1)$] provided one imposes some appropriate
boundary conditions.  For this reason, we call (\ref{I2}) the
Poincare
Chern-Simons action. The action (\ref{I2}) is a natural extension of
the
2+1 Chern-Simons gravity action studied in 
\cite{Achucarro-Townsend,Witten88} for zero cosmological constant.   

Our notations are the following:  $\eta_{AB}=diag(-1,1,1,1,\sigma)$;
Capital indices $A,B,...$ run over $SO(3,2)$ if $\sigma=-1$, and over
$SO(4,1)$ if $\sigma=1$.  $M$ is a five-dimensional
manifold with the topology $\Sigma\times \Re$ and $\Sigma$ has 
a boundary denoted by $\partial \Sigma$ (see Fig. 1). We 
shall call $B$ the ``cylinder" formed by the direct product of $\Re$ 
and $\partial \Sigma$,
\begin{equation}
B = \partial \Sigma \times \Re.
\end{equation}
According to Fig. 1, the surface $B$ is represented as the
hypersurface 
$x^5=const.$.  Local coordinates on $M$ are denoted by $x^M$,
$M=[0,1,2,3,4]$; local coordinates on $B$ are denoted by
$x^\mu$, $\mu=[0,1,2,3]$; and local coordinates of 
$\Sigma$, for all times, are denoted by $x^i$, $i =[1,2,3,4]$.  The
$SO(3,2)$ [$SO(4,1)$] covariant derivative is denoted by $\nabla$. \\

 \begin{center}



 Figure 1. \\

 $\Sigma_1$ and $\Sigma_2$ are the surfaces $t=t_1$ and
 $t=t_2$, respectively. \\
 $B=\partial \Sigma \times \Re$ is represented by $x^5=const.$\\
 \end{center}  


Varying (\ref{I2}) with respect to all the fields (assuming that the
boundary terms cancel out) we obtain the five-dimensional 
equations of motion
\begin{eqnarray}
\epsilon_{ABCDE} \tilde R^{AB}\mbox{\tiny 
$\wedge$} \tilde R^{CD} =0 \label{E},\\
\epsilon_{ABCDE} \tilde R^{AB}\mbox{\tiny 
$\wedge$} \tilde T^C =0, \label{W} 
\end{eqnarray}
where $\tilde T^A=\nabla e^A$ is the five-dimensional torsion 2-form.  
The configuration space defined by the above equations is stratified
into different regions with a different number of degrees of freedom. 
For
example, the configuration $\tilde R^{AB}=0=\tilde T^A$ solves the
above
equations but carries no local degrees of freedom. 
There exists, however, other
solutions for  which the curvatures do not vanish and represent
propagating modes.  The maximum number of (physical) local degrees of
freedom for  this theory is shown to be 13.  We refer the reader to 
\cite{BGH2} for details on this 
point. It is enough for our purposes here to
mention that on the ray of phase space that carries the maximum
number 
of
degrees of freedom, the above equations can be written in the useful  
form \cite{BGH2},
\begin{eqnarray}
\frac{d e^A_i}{dt} &=& D_i e^A_0 + N^k \tilde T^A_{ki}  \label{eq1}\\
\frac{d W^{AB}_i}{dt} &=& D_i W^{AB}_0 + N^k \tilde R^{AB}_{ki} 
\label{eq2}\\
0 &=& \epsilon^{ijkl} \epsilon_{ABCDE} 
            \tilde R^{AB}_{ij} \tilde R^{CD}_{kl} \label{c3} \\
0 &=& \epsilon^{ijkl} \epsilon_{ABCDE} 
            \tilde R^{AB}_{ij} \tilde T^{C}_{kl} \label{c4} 
\end{eqnarray}
where we have split $W^{AB}=W^{AB}_0 dt + W^{AB}_i dx^i$ and
$e^A=e^A_0
dt + e^A_i dx^i$.  Note that (\ref{eq1}) and (\ref{eq2}) explicitly
show
that the time evolution is generated by a  gauge transformation with
parameter $\{ W^{AB}_0,e^A_0\}$ plus a spatial (improved
\cite{Jackiw78}) diffeomorphism with parameter $N^k$.  As it was
shown
in \cite{BGH2} the normal deformations, or timelike diffeomorphisms,
do
not represent an independent symmetry. This is similar to what
happens
in 2+1 dimensions where the full diffeomorphism invariance is
contained
in the gauge group \cite{Witten88}. Note that the vector $N^k$ did
not
appear in the original Lagrangian. It appears here because in
obtaining
(\ref{eq1}) and (\ref{eq2}) from (\ref{E}) and (\ref{W}) a degenerate
matrix was  inverted. The vector $N^k$ parameterizes the linear
combination of null eigenvectors of that matrix \cite{BGH2}.    

To integrate (\ref{eq1}) and (\ref{eq2})  we only need to give Cauchy
data on the initial hypersurface. The initial data, on the other
hand,
must satisfy the constraints (\ref{c3}) and (\ref{c4}).  It is easy
to
see from (\ref{eq1}) and (\ref{eq2}) that the time evolution
preserves
the constraints, that is, a configuration satisfying (\ref{c3}) and
(\ref{c4}) at $t=t_1$, will remain on that surface at late times.
Eqs.
(\ref{eq1}) and (\ref{eq2}) can be studied, for example, in the gauge
$e^A_0=0$, $W^{AB}_0=0$ and $N^k=0$. In that gauge we find that
$e^A_i$
and $W^{AB}_i$ are time independent, thus, given their initial
values,
they are known for all times. \\

\noindent {\em 2. Boundary terms.} The above equations of motion
define an extremum for the action principle provided some suitable
boundary conditions are imposed such that all boundary terms cancel
out. The
boundary term coming from the variation of the action (\ref{I2}) is
easily computed  obtaining,    
\begin{equation}
-\int_{\partial M} \epsilon_{ABCDE} 
\tilde R^{AB}\mbox{\tiny $\wedge$} 
e^C \mbox{\tiny $\wedge$} \delta W^{DE}.
\label{BT}
\end{equation}
The boundary $\partial M$ has three connected components (see Fig.
1),
\begin{equation}
\partial M = \Sigma_1 \cup \Sigma_2 \cup B. 
\end{equation}
On the initial and final boundaries (denoted by $\Sigma_1$ and
$\Sigma_2$ in Fig. 1),  a natural way to cancel (\ref{BT}) is by
imposing $\delta W^{AB}_i=0$.  Of course, the possible values - or
Cauchy data - for $W^{AB}_i$ at $t=t_1$ and $t=t_2$ must satisfy the
constraints (\ref{c3}) and (\ref{c4}). Also, due to the first order
character of the equation for $W^{AB}_i$, its value on the final
surface
is --classically, but not  quantum mechanically-- related to the
data given on the initial surface. This is similar to what happens in
the example of a free particle in the momentum representation. Since
the
momentum is conserved, classically one cannot prescribe two different
values for $p(t)$ at $t=t_1$ and $t=t_2$, otherwise there will be no
solutions. Quantum mechanically, however, $p_1=p(t_1)$ and
$p_2=p(t_2)$ 
are completely independent but the propagation amplitude for going
from
$p_1$ to $p_2$ is shown to be proportional to $\delta(p_1-p_2)$. 
  
There is another boundary term appearing at a fixed value of $x^5$
for
all $t$ (denoted by $B$ in fig. 1).  On that surface, $W^{AB}$ cannot
be
fixed to an arbitrary value because that would over determine the
variational principle. Moreover, the action must define a propagation
amplitude (through the path integral) from an initial to a final
surface, without needing to prescribe data on the intermediate
states.  

There are many different ways to proceed in this situation. Perhaps
the
most natural way to treat this boundary term is by imposing
appropriated
boundary conditions such that (\ref{BT}) can be written in the form
$\delta X$, then one subtracts $X$ from the initial action thus
producing a new action whose variation does not generate any boundary
terms\cite{Regge-Teitelboim}. We shall carry out this procedure
below.  

Another method to deal with boundary terms would be to pass to the
Hamiltonian formalism. The Hamiltonian methods developed in
\cite{BGH2}
appear to be particularly appropriate to handle this case because a
$U(1)$ field  --necessary to apply the results of \cite{BGH2}-- can
be
coupled in a natural way to the action (\ref{I2}) \cite{B-Tr-Z}.
Thus,
one expects to find a four-dimensional current algebra at $B$, with a
Kahler form equal to the $U(1)$ field strength \cite{BGH2}. The
presence
of a current algebra at $B$ opens the possibility to find a
statistical
mechanical description \cite{Carlip95} for the entropy of the
five-dimensional black holes found in \cite{BTZ2}. We shall not
attack
this problem in this paper.

Finally, a third possibility to cancel (\ref{BT}) is simply to impose 
the coefficient of $\delta W^{AB}$ in (\ref{BT}) to be zero.  Note
that
this is similar to the open string (Newmann) boundary conditions
where
$\partial_\sigma x^\mu$ is set equal to zero in order to cancel the
boundary terms at the spatial boundaries. In this paper we shall
mainly
consider this latter possibility and study the resulting theory at
$B$. 

We thus impose the field equation at $B$ 
\begin{equation}
\epsilon_{ABCDE} \tilde R^{AB}\mbox{\tiny $\wedge$} e^C =0.  
\label{BT=0}
\end{equation}
Note that, in order to simplify the notation, we have used the same
symbols to represent the five-dimensional forms and their 
four-dimensional projections. The above equation is a
four-dimensional 
equation with $\tilde R^{AB}=\frac{1}{2}  \tilde R^{AB}_{\ \mu\nu}
dx^\mu\mbox{\tiny $\wedge$} dx^\nu$ and $e^A = e^A_\mu dx^\mu $ where
$x^\mu$, $(\mu=[0,1,2,3])$ are local coordinates on $B$. In terms of
local coordinates Eq. (\ref{BT=0}) reads
\begin{equation}
\epsilon^{\mu\nu\lambda\rho}\epsilon_{ABCDE} \tilde 
R^{AB}_{\ \mu\nu} e^C_\lambda=0.
\label{BT=0'}
\end{equation}  

Before going into the analysis of the consequences of (\ref{BT=0'})
it is necessary to check whether this equation completely defines the
dynamics of the fields at $B$. In other words, we would like to know
if the projections of (\ref{E}) and (\ref{W}) to $B$ become
identities after (\ref{BT=0'}) is imposed.  It turns out that
(\ref{W}) (projected
to $B$) is indeed satisfied as a consequence of (\ref{BT=0'}), but
the
same is not true for Eq. (\ref{E}).  The projections of (\ref{E}) and
(\ref{W}) to $B$ are, respectively,
\begin{eqnarray}
\epsilon^{\mu\nu\lambda\rho}\epsilon_{ABCDE} \tilde R^{AB}_{\ \mu\nu} 
\tilde R ^{CD}_{\ \lambda\rho}&=&0, 
\label{EB} \\
\epsilon^{\mu\nu\lambda\rho}\epsilon_{ABCDE} \tilde R^{AB}_{\ \mu\nu} 
\tilde T^C_{\ \lambda\rho}&=&0.
\label{WB}
\end{eqnarray}

Taking the $\nabla_\rho$ covariant derivative (tangential to $B$) of
Eq.
(\ref{BT=0'}) and using the Bianchi identity 
$\epsilon^{\mu\nu\lambda\rho} \nabla_\lambda \tilde R^{AB}_{\
\mu\nu}=0$, one obtains (\ref{WB}). Thus, (\ref{WB}) is indeed
identically satisfied once the boundary condition (\ref{BT=0'}) is
imposed.  

The situation is not so simple with Eq. (\ref{EB}). Eq. (\ref{EB}) is
quadratic in the curvature while (\ref{BT=0'}) is linear.  To see
whether (\ref{EB}) is a consequence of (\ref{BT=0'}) we multiply
(\ref{BT=0'}) by $R^{CD}_{\rho\sigma}$. Using some simple
combinatorial
identities  one obtains the equation,
\begin{equation}
\epsilon^{\mu\nu\lambda\rho}\epsilon_{ABCDE} R^{AB}_{\mu\nu} 
R ^{CD}_{\lambda\rho} \ e^E_\sigma=0.
\label{ee}
\end{equation}
This equation is almost what we need. Indeed, the coefficient of
$e^E_\sigma$ in (\ref{ee}) is exactly Eq. (\ref{EB}), but one
cannot infer from (\ref{ee}) that (\ref{EB}) holds because
$e^E_\sigma$
is not a squared matrix and thus it cannot be inverted.  However, the  
five-dimensional veilbein $e^A_M$ is invertible and therefore
$e^A_\mu$
has rank four.  This means that (\ref{ee})
does imply the vanishing of four of the five equations (\ref{EB}). To
isolate the part of (\ref{EB}) not contained in (\ref{ee}) we
need to break the explicit five-dimensional covariance.

We decompose the  five-dimensional indices $A,B,...$
into four-dimensional indices $a,b,...$ [a=(0,1,2,3)] and write  
$e^A_\mu$ and $W^{AB}_\mu$ in the form,
\begin{equation}
e^A_\mu=(e^a_\mu,e^5_\mu), \mbox{\hspace{1cm}} 
W^{AB}_\mu=(w^{ab}_\mu,W^{a5}_\mu). 
\label{1'}
\end{equation}
Note that under this decomposition all the
fields transform in a definite way under the Lorentz group $SO(3,1)$.
Indeed,  both, $e^a$ and $W^{a5}$ transform as vectors under
$SO(3,1)$;
$e^5$ is a Lorentz scalar and $w^{ab}$ is a Lorentz connection. We
can thus already identify the elements of general relativity, namely,
the tetrad $e^a$ and the spin connection $w^{ab}$. In this notation,
Eq. (\ref{ee}) takes the form 
\begin{equation}
\epsilon_{abcd}\epsilon^{\mu\nu\lambda\rho}
[\tilde R^{ab}_{\ \mu\nu} \tilde R^{cd}_{\ \lambda\rho}\ e^5_\sigma  
- 2\tilde R^{ab}_{\ \mu\nu}  
\tilde R^{c5}_{\ \lambda\rho}\ e^d_\sigma] =0.
\label{eee}
\end{equation}
If we borrow from (\ref{EB}) the single equation
$\epsilon_{abcd}\epsilon^{\mu\nu\lambda\rho}  \tilde R^{ab}_{\mu\nu}
\tilde R^{cd}_{\lambda\rho}=0$, then   invertibility of $e^d_\sigma$
together with (\ref{eee}) implies $\epsilon_{abcd}
\epsilon^{\mu\nu\lambda\rho} \tilde R^{ab}_{\mu\nu} \tilde
R^{c5}_{\lambda\rho}=0$.  We have thus isolated the part of Eq.
(\ref{EB}) not contained in (\ref{BT=0}), namely the equation
\begin{equation}
\epsilon_{abcd}\epsilon^{\mu\nu\lambda\rho} \tilde R^{ab}_{\mu\nu}
\tilde R^{cd}_{\lambda\rho}=0.
\label{E5}
\end{equation}
This equation,  contributes in a non-trivial way to the dynamics at
$B$.  

The independent field equations at $B$ are thus Eq. (\ref{BT=0}) plus
Eq. (\ref{E5}).  Now, one would like to know whether these equations
define a sensible field theory at $B$. One faces an immediate problem
because the number of equations and number of fields at $B$ does not
match. Indeed, the number of fields is 60 [$\#\{W^{AB}_\mu\}=40$ plus
$\#\{e^A_\mu\}=20$] while the number of equations is 41 [there are 40
equations in (\ref{BT=0}) plus Eq. (\ref{E5})]. One may wonder
whether
the remaining 19 fields may be a signal of some extra gauge symmetry.
This is not the case because we have counted here all the equations
including constraints. If there was some extra gauge symmetries we
should also find their associated constraints. [The number of
dynamical
fields plus Lagrange multipliers must equal the number of dynamical
equations plus constraints.] A second possible interpretation for
those
fields is as matter fields. This is certainly an interesting
possibility. Indeed,  in Kaluza-Klein theories this is exactly the
mechanism by which matter is brought in. In this particular case,
however, we have found no natural interpretation in that sense.       

To remedy the mismatch of fields and equations we purpose to 
impose further conditions on the fields at $B$.  We shall not attempt
to give a general set of extra possible boundary conditions. Rather,
we shall exhibit a set of conditions which
leave the right number of independent fields and the resulting theory
is dilaton gravity with a Gauss-Bonnet interaction. We impose at $B$
the
19 restrictions,  
\begin{equation}
e^5_\mu=l \partial_\mu \varphi \mbox{\hspace{1cm}} \mbox{and} 
\mbox{\hspace{1cm}} W^a_{\ 5\mu} = \frac{e^a_\mu}{l},
\label{1}
\end{equation}
where $\varphi$ is a dimensionless Lorentz scalar and 
$l$ is an arbitrary constant with dimensions of length. The
independent fields are then reduced to the four-dimensional tetrad
$e^a_\mu$, the spin connection $w^{ab}_\mu$ and --the dilaton--
$\varphi$.  The number of independent fields is thus 41, as required.   
Note that since $e^5$ is a Lorentz scalar and both, $e^a$ and
$W^a_{5}$ are Lorentz vectors, conditions (\ref{1}) are Lorentz
invariant. The local symmetry group of the boundary conditions 
has thus be broken down to the Lorentz group.

Some useful identities that hold under (\ref{1}) are 
\begin{eqnarray}
\tilde R^{ab} &=& R^{ab} - \frac{\sigma}{l^2} e^a 
               \mbox{\tiny $\wedge$} e^b, \label{Rtilde}\\ 
\tilde R^a_{\ 5} &=& \frac{T^a}{l}, \label{1a}\\ 
\tilde T^a &=& T^a + e^a \mbox{\tiny $\wedge$} d\varphi, \label{1b}
\\
\tilde T^5&=&0 \label{1c}
\end{eqnarray}
where $D$ is the $SO(3,1)$ covariant derivative, 
$R^{ab}$ the $SO(3,1)$
curvature two-form and $T^a=De^a$ the four-dimensional torsion. 
We also recall that $\sigma$ is the signature of the fifth dimension 
of the local group: $\eta_{AB}=diag(-1,1,1,1,\sigma)$. Hereafter 
we shall use the notation of differential forms. 
It is understood that they are forms defined at $B$.  

Replacing (\ref{1}) into (\ref{BT=0}) and using the above 
identities one 
obtains the $SO(3,1)$ equations 
\begin{eqnarray}
\epsilon_{abcd} \tilde R^{ab} \mbox{\tiny $\wedge$} e^c 
&=& 0, \label{ex1}\\
\epsilon_{abcd} (2\sigma T^a \mbox{\tiny $\wedge$} e^b 
- l^2 \tilde R^{ab} \mbox{\tiny $\wedge$} d\varphi) &=& 0,
              \label{ex2}
\end{eqnarray}
plus the equation (\ref{E5}) which reads,
\begin{equation}
\epsilon_{abcd} \tilde R^{ab} 
\mbox{\tiny $\wedge$} \tilde R^{cd} = 0. \label{c}
\end{equation}
These equations define the dynamics of the fields at $B$.  
\\

\noindent{\em 3. The effective field theory at $B$.}  
Equations (\ref{ex1}), 
(\ref{ex2}) and (\ref{c}) can be derived from the remarkable simple
four-dimensional action principle, 
\begin{equation}
I_B = k \int_B e^{\varphi} \epsilon_{abcd} 
\tilde R^{ab} \mbox{\tiny $\wedge$} \tilde R^{cd} 
\label{4D-action}
\end{equation}
where $\tilde R^{ab}$ was defined in (\ref{Rtilde}).  
It is direct to see that the variation with respect to `geometric'
variables $e^a$ and $w^{ab}$ give rise to Eqs. (\ref{ex1}) and
(\ref{ex2}), while the variation with respect to the dilaton
$\varphi$
give rise to (\ref{c}). $k$ is a coupling constant with dimension 
$(length)^2$. 

The action (\ref{4D-action}) can be thought of as a dilaton-like
generalization of the Mac-Dowell and Mansouri\cite{MacDowell} action. 
Note, however, that due to the presence of the dilaton, the
Gauss-Bonnet
term in (\ref{4D-action}) does contribute to the equations of motion. 
This is similar to what happens in 1+1 dimensions where the dilaton
field is included in order to provide a non-trivial dynamics  for the
two-dimensional Euler density. As a consequence, the 
variation of $I_B$ with respect to the spin connection [Eq.
(\ref{ex2})]
does not imply the vanishing of the torsion tensor, rather it gives a
$differential$ equation for the spin connection which, probably,
means 
that this theory has a dynamical torsion. A thorough analysis of
the solutions of (\ref{ex2}) is beyond the scope of this work. We
only
mention here that there exists a sector in the theory, for which the
torsion is equal to zero. That sector is simply obtained by
considering
the particular class of solutions for which the dilaton is constant,
\begin{equation}
\varphi=\varphi_0 
\label{phi=0}
\end{equation}
which together with Eq. (\ref{ex2}) imply $T^a=0$. We shall see below
that matter can be consistently coupled to the four-dimensional
theory
--keeping the 5D dynamics unaltered-- only on this sector of phase
space.  

The action (\ref{4D-action}) is written in the `string frame' where
the dilaton multiplies the whole action. By a conformal
transformation, one can pass to the `Einstein frame' in which the
Hilbert term decouples from the dilaton.  In this transformation
there exists an ambiguity due to the first order character of our
formalism. Since in this theory the tetrad and spin connection are
varied independently, one can make independent fields redefinitions
of them. The variation of the spin connection $w^{ab}$  under the
conformal transformation, 
\begin{equation}
e^a \rightarrow \Omega e^a
\end{equation}
is not dictated by the theory.  Of course one could use the
transformation dictated by the second order theory. If one does so
the action (\ref{4D-action}) is mapped into an action with
complicated couplings between the dilaton and curvature. The precise
formula for the transformed action can be
obtained by direct replacement of the formulas found, for example, in
\cite{Wald84} for conformal transformations.        

Perhaps a more interesting conformal transformation\footnote{We thank
C. Mart\'{\i}nez for help on this point.} is obtained by
simply rescaling the tetrad and leaving the spin connection
invariant.
If we redefine $e^a \rightarrow e^{-\varphi/2} e^a$ then the action 
(\ref{4D-action}) with  $k=-\sigma l^2/2$ is mapped into
\begin{equation}
I'_B=\int_B \epsilon_{abcd} (R^{ab} \mbox{\tiny $\wedge$} 
            e^c\mbox{\tiny $\wedge$} e^d - 
\frac{\sigma}{2l^2e^{\varphi}} e^a \mbox{\tiny $\wedge$} 
            e^b\mbox{\tiny $\wedge$} e^c\mbox{\tiny $\wedge$} e^d - 
\frac{\sigma l^2 e^{\varphi}}{2} R^{ab}\mbox{\tiny $\wedge$} R^{cd}).
\label{I4'}
\end{equation}
One can recognize in this action the Hilbert term, the cosmological
constant term (coupled to the dilaton) and the Gauss-Bonnet term also
coupled to the dilaton.  Since we did not transform the spin
connection, (\ref{I4'}) has no kinetical term for the dilaton. The
action (\ref{I4'}) --with a kinetical term for the dilaton-- appears
in
the 1-loop effective action of heterotic string theory and recently  
black holes solutions have been studied in \cite{Gauss-Bonnet/black}. 

The action (\ref{I4'}) has another interesting property:  it allows
the
elimination of  the dilaton from its own equation of motion. Indeed,
varying (\ref{I4'}) with respect to $\varphi$ one obtains an
$algebraic$
equation which can be solved for $\varphi$,
\begin{equation}
e^{-\varphi} = \frac{l^2}{2} \sqrt{\frac{\epsilon^{\mu\nu\lambda\rho}
\epsilon_{abcd} R^{ab}_{\mu\nu} R^{cd}_{\lambda\rho} }{24\ det(e)} }
\end{equation}
where $det(e)$ is the determinant of the tetrad. 
Replacing back this value of the dilaton into the action one 
obtains the rather curious modification of general relativity
\begin{equation}
I_{red} = \frac{1}{2}\int_B [\epsilon^{\mu\nu\lambda\rho}
\epsilon_{abcd} R^{ab}_{\mu\nu} e^c_\lambda  e^d_\rho + 2\sigma   
\sqrt{6\, det(e)\ \epsilon^{\mu\nu\lambda\rho}\epsilon_{abcd}
R^{ab}_{\mu\nu} R^{cd}_{\lambda\rho} }\ ]dx^4. 
\label{GR-GB}
\end{equation}
Note that the first term in (\ref{GR-GB}) is the usual
Einstein-Hilbert Lagrangian while the second term is the squared root
of the Gauss-Bonnet density. Note also that the second term
transforms correctly because both $det(e)$ and
$\epsilon^{\mu\nu\lambda\rho}$ are densities. 

Despite the fact that (\ref{GR-GB}) is non-linear in the curvature
and it is not even polynomial, it still gives rise to first order  
equations for the spin connection and tetrad.  Indeed, the equations
derived from (\ref{GR-GB}) are totally equivalent to the equations
derived from (\ref{I4'}).  In general, any function of the curvature
and
spin connection will give rise to first order equations for $w^{ab}$
and
$e^a$ provides one uses the Palatini formalism.  The price to pay is
that the equations of motion will not imply the vanishing of the
torsion
tensor. Indeed, in non-linear theories the Palatini and second order
formalisms are not equivalent.    \\

\noindent {\em 4. Coupling matter to the 4D theory.} 
The structure of the five-dimensional equations of motion (\ref{E})
and
(\ref{W}) suggest a natural way to couple a four-dimensional matter
Lagrangian to the dilatonic gravity at $B$. 
By virtue of the Bianchi identity, $\nabla \tilde R^{AB}=0$, Eq.
(\ref{W}) can be trivially integrated once obtaining,
\begin{equation}
\epsilon_{ABCDE} \tilde R^{AB} \mbox{\tiny $\wedge$} e^C = S_{DE}
\label{S}
\end{equation}
where $S_{DE}$ is a 3-form integration function satisfying 
\begin{equation}
\nabla S_{AB}=0 .
\label{con}
\end{equation}
[Consistency of (\ref{S}) with (\ref{E}) gives 
another condition for $S_{AB}$ that will be exhibited below.]

The 3-form $S_{AB}$ is an ``integration constant"  (covariantly
constant) of the five-dimensional equations of motion  and it
represents
part of the degrees of freedom of the theory. When projected to $B$,
(\ref{S}) is a four-dimensional  equation which looks like the
Einstein
equations with a non-zero energy momentum tensor $S_{AB}$, and
(\ref{con}) its  associated matter conservation equation. However,
under
our boundary conditions [see Eq. (\ref{BT=0})], we see that  the
projection
of $S_{AB}$ to $B$ has been imposed to vanish.  In order to be able
to
prescribe a non-zero value for $S_{AB}$ at $B$ --and thus find
dilaton
gravity with matter--  we add a boundary term to the 5D action.
Consider the action  
\begin{equation}
I = \frac{1}{2} \int_M \epsilon_{ABCDE} 
\tilde R^{AB}\mbox{\tiny $\wedge$} 
\tilde R^{CD}\mbox{\tiny $\wedge$} e^E  + \int_{B} L^m
(W^{AB},\phi) 
\label{I3}
\end{equation}
where $L^m(W,\phi)$ --the 4D matter Lagrangian-- is a given
functional 
defined at $B$. The matter fields will be collectively denoted by
$\phi$. Note that we have coupled the matter fields only to $W^{AB}$
and
not to the vielbein.  The reason is that $e^A$ enters without
derivatives in the five-dimensional action and therefore its
variation
does not couple to the fields at $B$.  Note however that after
imposing
(\ref{1}) at $B$, $W^{AB}$ contains the four-dimensional tetrad
through
$W^a_{\ 5}=e^a/l$.   

We vary the action (\ref{I3}) with respect to all the fields.  All
the
volume equations of motion remain the same  because we have only
added a
boundary term to the action. The only equation which is modified is
the
boundary equation (\ref{BT=0'}) which now reads
\begin{equation}
\epsilon_{ABCDE}\epsilon^{\mu\nu\lambda\rho}
 \tilde R^{AB}_{\mu\nu} e^C_\lambda = S^\rho_{DE},
\label{E2}
\end{equation}
where the ``energy momentum" tensor is defined by
\begin{equation}
S^\mu_{AB} = -\frac{\partial L^m(W,\phi)}{\partial W^{AB}_\mu}.
\end{equation}

The equations of motion for the matter follow from the  variation of
$L^m$ with respect to $\phi$. Note that since the boundary of $B$ are
the two disconnected ``rings" defined by the intersection of
$\Sigma_1$
and $\Sigma_2$ with $B$ [see Fig. 1], it is enough to prescribe
initial
and final values for the matter fields in order cancel all boundary
terms coming from the variation of the matter Lagrangian.  

We now study the compatibility of (\ref{E2}) with the bulk 
equations of motion projected to $B$, namely Eqs. (\ref{EB}) and
(\ref{WB}).  Compatibility of (\ref{E2}) with (\ref{WB}) impose 
\begin{equation}
\nabla_\mu S^\mu_{AB}=0,
\label{nablaS=0}
\end{equation} 
while compatibility of (\ref{E2}) with (\ref{EB}) impose
\begin{equation}
R^{AB}_{\ \mu\nu} S^\nu_{AB}=0.
\label{diff-con}
\end{equation}
Condition (\ref{nablaS=0}) follows directly from taking the
derivative of (\ref{E2}) and comparing with (\ref{WB}). Condition 
(\ref{diff-con}) is obtained by multiplying (\ref{E2}) by
$R^{DE}_{\rho\sigma}$, using some simple combinatorial identities
and comparing with (\ref{EB}).
The geometrical meaning of Eqs. (\ref{nablaS=0}) and (\ref{diff-con})
is
straightforward and it provides a remarkable link between the 
five-dimensional equations of motion and the symmetries of the 
four-dimensional matter Lagrangian. Indeed, 
Eq. (\ref{nablaS=0}) is associated to invariance of 
$L^m(W,\phi)$ under local gauge transformations 
\begin{equation}
\delta_\lambda W^{AB}_\mu = -\nabla_\mu \lambda^{AB}
\label{gauge}
\end{equation} 
where $\lambda^{AB}$ is an arbitrary parameter.  Eq.
(\ref{diff-con}), 
on
the other hand, is associated to invariance of $L^m(W,\phi)$ under  
(improved \cite{Jackiw78}) diffeomorphisms 
\begin{equation}
\delta_\xi W^{AB}_\mu = R^{AB}_{\ \mu\nu} \xi^\nu
\end{equation} 
where $\xi^\mu$ is an arbitrary vector field at $B$. 
Thus, compatibility between the boundary condition (\ref{E2}) and the
five-dimensional equations of motion is ensured provided the matter
Lagrangian is invariant under local gauge transformations and
diffeomorphisms. 

Invariance under diffeomorphisms is the minimum symmetry that any
Lagrangian coupled to gravity should possess. Hence, Eq. 
(\ref{diff-con}) will be satisfied by any reasonable choice of $L^m$.
Actually, we shall see shortly that for torsionless configurations,
Eq.
(\ref{diff-con}) is identically satisfied. That is the reason that
(\ref{diff-con}) is not commonly found in the literature.

The invariance of $L^m$ under the gauge transformations (\ref{gauge})
is
more subtle. The gauge transformation (\ref{gauge}) is not a simple
Lorentz rotation, rather it implies a full local  five-dimensional
$SO(3,2)$ symmetry [or SO(4,1) depending on the sign of $\sigma$].
Even
though one can construct some toy examples of matter Lagrangians
possessing such symmetry, most of the phenomenological and commonly
used
forms of matter are not $SO(3,2)$ invariant but only invariant under
the
subgroup of Lorentz transformations $SO(3,1)$.  The problem faced
here
can be put in a different, perhaps more transparent, way.  As we saw
in
the vacuum case, in order to find a sensible theory at $B$ we need to
impose the conditions  (\ref{1}). The equations derived from
(\ref{E2}) after imposing (\ref{1}) are
\begin{eqnarray}
\epsilon_{abcd} \tilde R^{ab} \mbox{\tiny $\wedge$} e^c 
&=& S_d,  \label{ex1m}\\
\epsilon_{abcd} (2\sigma T^a \mbox{\tiny $\wedge$} e^b 
- l^2 \tilde R^{ab} \mbox{\tiny $\wedge$} d\varphi) &=& l S_{cd}, 
              \label{ex2m}
\end{eqnarray}
where we have defined 
\begin{equation}
S_{a} = - \sigma l \frac{\partial L^m}{\partial e^a},
\mbox{\hspace{1cm}}
S_{ab} = - \frac{\partial L^m}{\partial w^{ab}}. 
\end{equation}
Note that these equations are just the generalizations of (\ref{ex1}) 
and (\ref{ex2}) with matter. As in the vacuum case, the fifth 
component of (\ref{EB}) contributes in a non-trivial way 
to the dynamics at $B$. Hence we add the equation,
\begin{equation}
\epsilon_{abcd} \tilde R^{ab}\mbox{\tiny $\wedge$} \tilde R^{cd} =0.
\label{c'}
\end{equation} 
which, together with (\ref{ex1m}) and (\ref{ex2m}), define the field
theory at $B$. 
    
Eqs. (\ref{ex1m}) and (\ref{ex2m}) are Lorentz and diffeomorphism
invariant, but they do not possess the larger $SO(3,2)$ [$SO(4,1)$]
local symmetry.  Therefore the symmetries of the left hand side of
(\ref{ex1m}) and (\ref{ex2m}) are not enough to ensure
(\ref{nablaS=0}).
There exists, however, a large class of solutions of the equations of
motion which do satisfy (\ref{nablaS=0}) and (\ref{diff-con}).
Consider
the case of spinless matter ($S_{ab}=0$) and solutions for which the
torsion  tensor is equal to zero $(T^a=0)$. Under this restriction
both, (\ref{nablaS=0}) and (\ref{diff-con}) are identically
satisfied. 
Indeed, (\ref{diff-con}) is equivalent to 
\begin{equation}
\tilde R^{ab}_{\ \mu\nu} S^\nu_{ab} + 2 T^a_{\ \mu\nu} S^\nu_a =0
\end{equation}
which is identically satisfied for 
$T^a=0=S_{ab}$.  On the other hand, Eq. (\ref{nablaS=0}) written in
the
$SO(3,1)$ language becomes
\begin{eqnarray}
(\sigma l) D_\mu S^\mu_{\ a} + S^\mu_{\ ab} e^b_\mu =0, \label{T1}\\
(\sigma l) D_\mu S^\mu_{\ ab} - e_{\mu a} S^\mu_{\ b} + e_{\mu b}
S^\mu_{\ a}=0.
\label{T2}
\end{eqnarray}
For $S^{ab}=0$, these equations represent diffeomorphism and Lorentz 
invariance of (\ref{ex1m}) and (\ref{ex2m}). Indeed, 
(\ref{T1}) is equivalent to the conservation of the energy-momentum
tensor $T^{\mu\nu} = S^{\mu a} e^\nu_a$ while (\ref{T2}) imposes
$T^{\mu\nu}=T^{\nu\mu}$.    

A different point of view for the same issue is provided by the study
of
the transformation for the tetrad induced by (\ref{gauge}): 
\begin{equation} 
\delta
e^a_\mu = -D_\mu \lambda^a + \lambda^a_{\ b} e^b_\mu  
\label{de} 
\end{equation}
where $\lambda^{AB} = (\lambda^{ab}, (\sigma/l) \lambda^a )$.   The
second term in (\ref{de}) is just a local Lorentz rotation which is
in
fact a symmetry of the equations of motion (\ref{ex1m}) and
(\ref{ex2m}). The first term is a symmetry of the equations only in
the
case when the torsion tensor is zero. In that case, $\delta e^a_\mu =
-D_\mu \lambda^a$ represents a diffeomorphism with a parameter
$\xi^\mu
=e^\mu_a \lambda^a$.  Note that for spinless matter ($S_{ab}=0$) the
matter Lagrangian is  independent of the spin connection and
therefore
the transformation induced by (\ref{gauge}) is irrelevant. 
 
In summary, we have seen that the dynamics at the boundary can be
altered by the adding of a boundary matter Lagrangian without
modifying
the 5D equations only in the sector for which the torsion is zero.
It
is rather disturbing that we have been forced to set the torsion
tensor
equal to zero at $B$ rather than deriving it from the equations of
motion. The relevant question --which will not be addressed here-- is
whether setting $T^a=0$ kills any degrees of freedom. A less
harmful way to set $T^a=0$ is by considering the class of solutions
for
which the dilaton is constant.  For a constant dilaton --which can
thus
be identified with Newton's constant-- Eqs. (\ref{ex1m}) and
(\ref{ex2m}) are exactly the Einstein equations with matter and for
$S_{ab}=0$ Eq. (\ref{ex2m}) implies $T^a=0$.  \\

\noindent {\em 5. Anti-de Sitter Chern-Simons gravity}.  The reader
familiar with CS gravity may wonder if the above considerations can
be applied to the (anti-) de Sitter CS theory. The answer to that
question is affirmative for (and only for) the anti-de Sitter 
theory as we now explain.  

The (anti-) de Sitter CS theory in five dimensions is defined 
by the action\cite{Chamseddine,BTZ2},
\begin{equation}
I_L = \frac{1}{2}\int \epsilon_{ABCDE} [\tilde R^{AB} 
\mbox{\tiny $\wedge$} \tilde R^{CD} \mbox{\tiny $\wedge$} e^E - 
\frac{2\kappa}{3L^2} \tilde R^{AB}\mbox{\tiny $\wedge$} 
e^C\mbox{\tiny $\wedge$} e^D\mbox{\tiny $\wedge$} e^E  
+\frac{1}{5L^4} e^A\mbox{\tiny $\wedge$} e^B \mbox{\tiny 
$\wedge$} e^C\mbox{\tiny $\wedge$} e^D\mbox{\tiny $\wedge$} e^E ]
\label{Ids}
\end{equation}
where, as before, $\eta_{AB}=diag(-1,1,1,1,\sigma)$ and $\kappa=\pm
1$. Depending on the signs of $\sigma$ and $\kappa$ the action
(\ref{Ids}) can be shown to be a CS action for the
groups\cite{Chamseddine}:   
\begin{eqnarray}
SO(5,1) &\ \ \ if\ \ \ & \sigma=\kappa=1 \mbox{\hspace{1cm}} 
\mbox{(de Sitter)} \nonumber\\
SO(4,2) &\ \ \ if\ \ \ & \sigma=-\kappa  \mbox{\hspace{1cm}} 
\mbox{(anti-de Sitter)}
                                            \label{groups} \\
SO(3,3) &\ \ \ if\ \ \ & \sigma=\kappa=-1. \nonumber   
\end{eqnarray}
In (\ref{Ids}) the term linear in the curvature is the 5D Hilbert
term, 
while the last term is the five-dimensional cosmological constant, 
parameterized by $L$. 
The action (\ref{Ids}) with $\sigma=1$ has a sensible interpretation 
as a 5D gravitational action, and, in the case $\kappa=-1$ black
holes
solutions exist\cite{BTZ2}. Also, it can be proved that the above
action
does possess local degrees of freedom\cite{BGH2}.   

We shall now study the problem of boundary terms and boundary
conditions
arising in the variation of (\ref{Ids}).  We vary (\ref{Ids})
with respect to all the fields obtaining the bulk equations of motion
\begin{eqnarray}
\epsilon_{ABCDE} F^{AB} \mbox{\tiny $\wedge$} F^{CD} =0
\label{Eds},\\
\epsilon_{ABCDE} F^{AB} \mbox{\tiny $\wedge$} \tilde T^A =0
\label{Wds},
\end{eqnarray}
where $\tilde T^A =\nabla e^A$ and we have defined
\begin{equation}
F^{AB} = \tilde R^{AB} - \frac{\kappa}{L^2} 
e^A\mbox{\tiny $\wedge$} e^B. 
\label{Rtilde'}
\end{equation}
We also obtain the equation at $B$
\begin{equation}
\epsilon_{ABCDE} ( F^{AB}\mbox{\tiny $\wedge$} e^C 
+ \frac{2\kappa}{3L^2}
e^A\mbox{\tiny $\wedge$} e^B\mbox{\tiny $\wedge$}
e^C)=0, 
\label{Bds}
\end{equation}
ensuring that all boundary terms at $B$ are equal to zero. We stress
that while Eq. (\ref{Eds}) and (\ref{Wds}) are five-dimensional
equations, Eq. (\ref{Bds}) is a four-dimensional equation defined
only
at $B$. It is worth noticing the similarity between (\ref{Eds}) and
(\ref{Wds}) with their Poincare counterparts Eqs. (\ref{E}) and
(\ref{W}). Eqs. (\ref{Eds}) and (\ref{Wds}) are obtained from
(\ref{E}) and (\ref{W}) just by replacing $\tilde R^{AB}$ by $
F^{AB}$. The equation at the boundary, on the other hand, does not
enjoy the same property [see (\ref{BT=0}) and (\ref{Bds})]. We shall
see shortly that the extra term appearing in (\ref{Bds}) is necessary
for compatibility between (\ref{Bds}) and the projections to $B$ of 
(\ref{Eds}) and (\ref{Wds}).

As we did in the Poincare theory, we need to study the compatibility
of
the boundary condition (\ref{Bds}) with the bulk equations of motion. 
The procedure is very much like in the Poincare theory, so we shall
only
quote the main results.  In the following, we only consider the
projections of (\ref{Eds}) and (\ref{Wds}) to $B$ which, in
differential
forms notation, look the same so we do not write them again. We first
note that Eq. (\ref{Wds}) (projected to $B$) is a direct consequence
of
(\ref{Bds}). Indeed, taking the covariant ($\nabla$) derivative of
(\ref{Bds}) we get (\ref{Wds}). Hence, any solution satisfying
(\ref{Bds}) will automatically satisfy (\ref{Wds}) at $B$.  Eq.
(\ref{Eds}) is more complicated. Just as in the Poincare case, only
part
of equation (\ref{Eds}) is satisfied once (\ref{Bds}) is imposed. 
Again, to explicitly isolate the part of (\ref{Eds}) not contained in
(\ref{Bds}), we break the explicit five-dimensional local symmetry as
in
(\ref{1'}) and then  we impose (\ref{1}) in order to have the same
number of fields and equations at $B$. A special feature of the
(anti-)
de Sitter theory is that the consistency of the boundary condition
(\ref{Bds}) with Eq. (\ref{Eds}) fixes the parameter $l$ appearing in
(\ref{1}) in terms of the parameter $L$ appearing in (\ref{Ids}) by
$l^2=L^2$. Also, the signatures $\sigma$ and $\kappa$ will be related
by
$\sigma=-\kappa$. Thus, according to (\ref{groups}) only the anti-de
Sitter [$SO(4,2)$] theory admits the boundary condition (\ref{Bds}).
We
note that the action (\ref{Ids}) admits black holes solutions only in
this case \cite{BTZ2}.     

A quick way to obtain this restrictions on the parameters $l$ and
$\sigma$ is by considering the equation (\ref{Eds}) [projected 
to $B$] in terms of its $SO(3,1)$ components 
\begin{eqnarray}
\epsilon_{abcd} F^{ab} \mbox{\tiny $\wedge$} F^{c5} =0 
\mbox{\hspace{1cm}} E=e, \label{Eds1} \\
\epsilon_{abcd} F^{ab} \mbox{\tiny $\wedge$} F^{cd} =0 
\mbox{\hspace{1cm}} E=5. \label{Eds2}
\end{eqnarray}
The compatibility of (\ref{Eds1}) with (\ref{Bds}) can be checked
using
(\ref{Wds}), which we already know is consistent with (\ref{Bds}).
Indeed, the component $E=5$ of (\ref{Wds}) reads,
\begin{equation}
\epsilon_{abcd} F^{ab} \mbox{\tiny $\wedge$} \tilde T^c =0.
\label{Wds1}
\end{equation}    
Therefore, comparing (\ref{Eds1}) with (\ref{Wds1}) we see that they
become identical if $F^{a5} = f\ \tilde T^a $ where $f$ is any smooth
function.  By direct application of (\ref{1a}), (\ref{1b}) and 
(\ref{Rtilde'}) we find that  $F^{a5} = f\ \tilde T^a $ 
implies, 
\begin{equation}
\frac{\sigma}{l} T^a- \frac{l\kappa}{L^2}e^a
\mbox{\tiny $\wedge$} d\varphi=
f [T^a + e^a  \mbox{\tiny $\wedge$} d\varphi]
\end{equation}
an equality that holds for arbitrary values of $T^a,e^a$ and
$\varphi$ 
only for $f=\sigma/l$ and 
\begin{equation}
\sigma =-\kappa, \mbox{\hspace{1cm}}  l^2 = L^2.
\label{l=L}
\end{equation}

In complete analogy with the 
Poincare case, Eq. (\ref{Eds2}) is not a consequence of the boundary
conditions (\ref{Bds}) and therefore it contributes in a non-trivial
way
to the dynamics at $B$. Thus, the theory at $B$ is described
by the equations (\ref{Bds}) plus (\ref{Eds2}). Note however that
condition (\ref{l=L}) imply that $F^{ab}=R^{ab}$ and thus
(\ref{Eds2}) becomes simply,
\begin{equation}
\epsilon_{abcd} R^{ab}\mbox{\tiny $\wedge$} R^{cd} =0.
\label{Eds22}
\end{equation} 

The equations at $B$ can again be derived from a simple
four-dimensional
action principle,
\begin{equation}
I = \int_B \epsilon_{abcd} ( R^{ab}\mbox{\tiny $\wedge$} 
e^c \mbox{\tiny $\wedge$} e^d
 + \frac{\kappa}{3L^2} e^a\mbox{\tiny 
$\wedge$} e^b\mbox{\tiny $\wedge$} e^c\mbox{\tiny $\wedge$} e^d -
\frac{\kappa L^2}{2} R^{ab}\mbox{\tiny $\wedge$} R^{cd} \varphi ) 
\label{IBds}
\end{equation}
in which the dilaton is coupled to the Gauss-Bonnet term only. Note
that
the first and second terms are the usual Einstein-Hilbert and
cosmological terms.  
 
The variation of (\ref{IBds}) with respect to $\varphi$ gives
(\ref{Eds22}), while the variation with respect to the ``geometrical"
variables $w^{ab}$ and $e^a$ gives, respectively,  
\begin{eqnarray}
\epsilon_{abcd} (2T^a \mbox{\tiny $\wedge$} e^b - 
\kappa L^2 R^{ab}\mbox{\tiny $\wedge$} d\varphi)=0 \\
\epsilon_{abcd} (R^{ab}\mbox{\tiny $\wedge$} e^c +
\frac{2\kappa}{3L^2} 
e^a\mbox{\tiny $\wedge$} e^b\mbox{\tiny $\wedge$} e^c)=0.
\end{eqnarray}
These equations are exactly equivalent to (\ref{Bds}) written in
$SO(3,1)$ components, after (\ref{1}) is imposed. The 
action (\ref{IBds}) can also be interpreted as a
dilaton-like gravity. Note that the coupling of the dilaton with the
gravitational variables is different to the Poincare case.  \\

\noindent {\em 6. Other boundary conditions.}  Finally, we would like
to
consider other possible boundary conditions.  Let us go back to Eq.
(\ref{BT}) and study a different method to treat the boundary terms,
namely, we shall now impose boundary conditions such that (\ref{BT})
can
be written as a total variation.  
 
The boundary term (\ref{BT}) is clearly not a total variation unless
some boundary conditions linking $e^A$ with $W^{AB}$ are imposed. 
One could also use boundary conditions for which $e^A$ is fixed at
$B$.
However, those boundary conditions may over-determine the Cauchy
problem
and also the propagating amplitudes would not depend on initial and
final data but also on intermediate data defined at $B$. To avoid the
need of giving data at $B$, the standard procedure is to
set some fields equal to a fixed, not arbitrary, given value (e.g.
asymptotic flatness in GR). 

We thus face the problem of imposing boundary conditions such  that
(\ref{BT}) can be written as a total variation. It is remarkable that
precisely (\ref{1}) provides such conditions for a fixed value of
$\varphi$. For simplicity, we fix $\varphi=\varphi_0$ which implies
$e^5=0$. It is easy to check that under (\ref{1}) one has the
identity
\begin{equation}
\int_{B} \epsilon_{ABCDE} \tilde 
R^{AB}\mbox{\tiny $\wedge$} e^C \mbox{\tiny $\wedge$} \delta W^{DE} = 
\delta \left[ \frac{\sigma}{2l} \int_B \epsilon_{abcd} (
2 R^{ab} -\frac{\sigma}{l^2} e^a\mbox{\tiny $\wedge$} e^b) 
\mbox{\tiny $\wedge$} e^c\mbox{\tiny $\wedge$} e^d
\right]
\label{deltaBT}
\end{equation}
[This identity involves an integration by parts in the right hand
side 
which generates a boundary term at $\partial B$. Since the fields at
$\partial B$ are fixed, this boundary term is equal to zero.] 

We thus consider the improved five-dimensional action 
\begin{equation}
I = \frac{1}{2}\int_M \epsilon_{ABCDE} 
\tilde R^{AB} \mbox{\tiny $\wedge$} \tilde R^{CD} 
\mbox{\tiny $\wedge$} e^E 
+ \frac{\sigma}{2l} \int_B \epsilon_{abcd} (
2 R^{ab} -\frac{\sigma}{l^2} e^a\mbox{\tiny $\wedge$} e^b) 
\mbox{\tiny $\wedge$} e^c\mbox{\tiny $\wedge$} e^d 
\label{action/f}
\end{equation}
which has well-defined functional derivatives under variations
satisfying condition (\ref{1}) at $B$.  Recall that now (\ref{BT=0})
is
not imposed. Note that the boundary term that we have added is just
the
Einstein-Hilbert action with a cosmological constant. 

The variation of the action (\ref{action/f}) give rise to equations
(\ref{W}) and (\ref{E}) which together with the boundary conditions
completely define the theory.  Since we have not imposed Eq.
(\ref{BT=0}) at $B$, we do not find  a ``boundary theory" in this
case. 
Despite the presence of the Einstein-Hilbert action in 
(\ref{action/f}), the equations of motion at $B$ are the projections
to
$B$ of (\ref{W}) and (\ref{E}) which are not equivalent to the
Einstein
equations.  However, one can speculate that quantum mechanically the
action (\ref{action/f}) induces general relativity at $B$, at least
in a saddle point approximation of the path integral.  This can be
seen
as follows. 

A key property of the CS action is that the value of the bulk term
evaluated on $any$ solution of the equation of motion is equal to
zero.
Therefore, the value of I given in (\ref{action/f}) evaluated on any
solution of the equations of motion is equal to the boundary term,
that
is, equal to the Einstein-Hilbert action at $B$. Since in the CS
variational principle we have only imposed (\ref{1}) with 
$\varphi=\varphi_0$, we have to vary the boundary action with respect
to
the remaining fields, that is, $e^a$ and $w^{ab}$ thus obtaining
general
relativity at $B$. 
 
\acknowledgements

The author would like to thank L.J. Garay, A. Gomberoff, M. Henneaux,
C. Mart\'{\i}nez and C. Teitelboim for many useful conversations.
This work was partially supported by a grant from Fundaci\'on Andes
(Chile), grants \# 1960065 and \# 1940203 from FONDECYT (Chile) and
institutional support by a group of Chilean companies (EMPRESAS CMPC,
CGE, COPEC, CODELCO, MINERA LA ESCONDIDA, NOVAGAS, ENERSIS, BUSINESS
DESIGN ASS. and XEROX Chile).

\end{document}